\documentclass[12pt]{article}
\usepackage{latexsym,graphicx}
\newcommand{\be}{\begin{equation}}
\newcommand{\ee}{\end{equation}}
\usepackage[left=2.5cm,top=2.5cm,right=2.5cm,bottom=1.5cm]{geometry}
\begin{document}
\begin{center}
\large{\bf{Bianchi type-I transit cosmological models with time dependent gravitational and cosmological constants$-$reexamined}} \\
\vspace{10mm}
\normalsize{{A Pradhan {\footnote{Corresponding author, Email: pradhan@iucaa.ernet.in; pradhan.anirudh@gmail.com}}}, 
B Saha$^{2}$, V Rikhvitsky$^{2}$}\\
\vspace{5mm}
\normalsize{$^{1}$Department of Mathematics, G. L. A. University, Mathura-281 406, Uttar Pradesh, India} \\
\vspace{5mm}
\normalsize{$^{2}$Laboratory of Information Technologies, Joint Institute for Nuclear Research, 141980 Dubna, 
Moscow Region, Russia} \\
\end{center}
\vspace{10mm}
\begin{abstract}
The present study reexamines the recent work of Pradhan et al. (Indian J. Phys. 88: 757, 2014) and obtained general 
exact solutions of the Einstein's field equations with variable gravitational and cosmological ``constants'' for 
a spatially homogeneous and anisotropic Bianchi type-I space-time. To study the transit behaviour of Universe, we 
consider a law of variation of scale factor $a(t) = \left(t^{k} e^{t}\right)^{\frac{1}{n}}$ which yields a time 
dependent deceleration parameter $q = - 1 + \frac{nk}{(k + t)^{2}}$, comprising a class of models that depicts a 
transition of the universe from the early decelerated phase to the recent accelerating phase. We find that the 
time dependent deceleration parameter is reasonable for the present day Universe and give an appropriate description 
of the evolution of the universe. For $n = 0.27k$, we obtain $q_{0} = -0.73$ which is similar to observed value of 
deceleration parameter at present epoch. It is also observed that for $n \geq 2$ and $k = 1$, we obtain a class of 
transit models of the universe from early decelerating to present accelerating phase. For $k = 0$, the universe has 
non-singular origin. In these models, we arrive at the decision that, from the structure of the field equations, the 
behaviour of cosmological and gravitational constants and are related. Taking into consideration the observational data, 
we conclude that the cosmological constant behaves as a positive decreasing function of time whereas gravitational 
constant is increasing and tend to a constant value at late time. $H(z)/(1+z)$ data ($32$ points) and model prediction 
as a function of redshift for different $k$ and $n$ are successfully presented by using recent data (Farooq and Ratra, 
Astrophys. J. 66: L7, 2013). Some physical and geometric properties of the models are also discussed.
\end{abstract}
\smallskip
PACS: 98.80.Es; 98.80.-k \\
Keywords : Cosmology; Transit universe; Variable gravitational \& cosmological constants; Variable deceleration parameter\\
\section{Introduction}
In Einsteinian field equations, the cosmological constant ($\Lambda$) and the gravitational constant ($G$) are two essential 
parameters. Accelerated expansion of the universe is due to dark energy as discovered by type Ia supernovae (SN Ia) 
observations as reported by Riess et al. \cite{ref1} and Perlmutter et al. \cite{ref2}. The Wilkinson Microwave Anisotropy 
Probe (WMAP) \cite{ref3,ref4}, combined with more accurate SN Ia data \cite{ref5} indicates that the Universe is almost spatially 
flat and the dark energy accounting to approximately to $70\%$ of the total content of the Universe. Little is known about 
the nature of dark energy. Observations strongly favour a small and positive value of the effective cosmological constant 
at the present epoch. Among many possible alternatives, the simplest and theoretically appealing possibility of dark energy 
is the energy density stored on the vacuum state of all existing fields in the universe i. e., $\rho_{v} = \frac{\Lambda}{8\pi G}$. 
The variable cosmological constant [6$-$8] is one of the phenomenological ways to explain the dark energy problem, compatible 
with observations. The problem in this approach is to determine the right dependence of $\Lambda$ upon scale factor $R$ or 
cosmic time $t$. Motivated by dimensional grounds with quantum cosmology, the variation of cosmological term as 
$\Lambda \propto R^{-2}$ is considered by Chen and Wu \cite{ref9}. However, several ans$\ddot{a}$tz have been proposed 
in which the $\Lambda$-term decays with time [10$-$13]. Several authors [13$-$18] have recently studied the time dependent 
cosmological constant in different contexts. Recently, Arkhipova et al. \cite{ref21} have studied growth rate in dynamical 
dark energy models by using a compilation of baryonic acoustic oscillation (BAO), growth rate and the distance prior from the 
cosmic microwave background (CMB) to constrain model parameter of the scalar field model. When combining these constraints 
with the constraints coming from a distance-redshift relationship (BAO data and the distance prior from CMB) the degeneracy 
is broken and one obtains $\Omega_{m} = 0.30 \pm 0.04$ and $\alpha < 1.30$ with the best fit value of $\alpha = 0.00$. 
Pavlov et al. \cite{ref22} compile a list of $14$ independent measurements of large-scale structure growth rate between 
redshift $0.067 \leq z \leq 0.8$ and use this to place constraints on model parameters of constant and time-evolving 
general-relativistic dark energy cosmologies. Chen et al. \cite{ref23} have recently pointed out that the recent observations 
still can not distinguish whether dark energy is a time-independent or a time-varying dynamical component through a joint 
analysis of the strong gravitational lensing data with the more restrictive updated Hubble parameter measurements and the 
type Ia supernova data from SCP Union2. Farooq, Mania and Ratra \cite{ref24} have recently constrained two non-flat time 
evolving dark energy cosmological models by using Hubble parameter data, Type Ia supernova apparent magnitude measurements, 
and baryonic acoustic oscillation peak length scale observations. These data require the average magnitude of the curvature 
density parameter $\mid\Omega_{k0}\mid$ is less or equivalent to $0.15$ at $1\sigma$ confidence. \\

Recent observations do not support the stability of fundamental constants and ``Equivalence Principle'' of general relativity. 
Dirac \cite{ref25,ref26} was the first to introduce the time variation of the gravitational constant $G$ in his large 
number hypothesis and after that it has been used in various modifications of general theory of relativity. The applications 
to observable tests, use of the recent techniques developed for handling the observational tests for $G$-varying cosmologies 
are discussed by Canuto et al. \cite{ref27,ref28}. Theories of gravity in which the $G$ varies with time and cosmological 
models based on them have been extensively studied in literature [29$-$31]. The variation of gravitational constant inferred 
from the Hubble diagram of Type Ia Supernovae \cite{ref30}. For the case of the gravity theory we can have varying-$G$, 
varying-$c$, or varying-$\Lambda$ models [32$-$34]. In these models either $G$ or $c$ is described by a scalar field and 
the equation of motion determine the dynamics of both the space-time metric and the varying constant. In the context of 
cosmological models, according to cosmological principle these varying constants are only time-dependent. Several authors 
[35$-$40] have recently investigated and discussed the time dependent $\Lambda$ and $G$ in different contexts. Recently, 
Yadav and Sharma \cite{ref41} and Yadav \cite{ref42} have discussed about transit universe in Bianchi type-V space-time 
with variable G and $\Lambda$.  \\

It is important to mention here that CMB anisotropy data are sometimes viewed as supporting a very small breaking of 
spatially isotropy. Recently, Ade et al. \cite{ref43} have investigated Planck 2013 results. According to their findings, 
deviations from isotropy have been found and demonstrated to be robust against component separation algorithm, mask choice 
and frequency dependence. In particular, they investigated that the quadrupole-octopole alignment is also connected to a 
low observed variance of the CMB signal. While we consider a Bianchi model here, another possible source of this anisotropy 
is a very small cosmological magnetic field \cite{ref44,ref45}. Chen et al. \cite{ref44} have discussed the constraints 
that can be placed on the strength of such a primordial magnetic field using CMB anisotropy data from WMAP experiments. 
Kahniashvili, Lavrelashvili and Ratra \cite{ref45} have indicated CMB temperature anisotropy from broken spatial isotropy 
due to homogeneous cosmological magnetic field. The analysis of WMAP data sets shows that the universe could have a 
preferred direction. The ILC-WMAP data maps show seven axes well aligned with one another and the direction Virgo. For this 
reason Bianchi models are important in the study of anisotropies. \\

It is well established that anisotropic Bianchi type-I universe plays significant role in understanding the phenomenal 
formation of galaxies in early universe. Theoretical reasoning and recent observations of cosmic microwave 
background radiation (CMBR) subscribe to the existence of anisotropic phase that approaches an isotropic one. Recently, 
Yadav et al. \cite{ref49,ref50} have obtained Bianchi type-I cosmological models with viscosity and cosmological term in 
general relativity by considering scale factor as $a(t) = \sqrt{(t e^{t})}$ and $a(t) = \sqrt{(t^{n} e^{t})}$ respectively. 
Pradhan et al. \cite{ref60} have also investigated accelerating Bianchi type-I universe with time varying $G$ and 
$\Lambda$-term in general relativity by taking $a(t) = \sinh^{\frac{1}{n}}(\alpha T)$. Recently, Pradhan et al. \cite{ref51} 
have studied Bianchi type-I transit cosmological models with time dependent gravitational and cosmological constants by 
considering scale factor as $a(t) = \sqrt{(t^{n} e^{t})}$. Motivated by the above discussions, in this paper, we propose to study 
homogeneous and anisotropic Bianchi type-I transit cosmological models with time dependent gravitational and cosmological 
``constants'' by assuming $a(t) = \left(t^{k} e^{t}\right)^{\frac{1}{n}}$ which generalize our previous studies and provide 
better results. 

\section{The metric and basic equations}
We consider the space-time metric of the spatially homogeneous and anisotropic Bianchi-I of the form
\begin{equation}
\label{eq1}
ds^{2} = dt^{2} - A^{2}(t) dx^{2} - B^{2}(t)dy^{2} - C^{2}(t)dz^{2}.
\end{equation}
where A(t), B(t) and C(t) are the metric functions of cosmic time t. \\

Einstein field equations with time-dependent $G$ and $\Lambda$ are given by
\begin{equation}
\label{eq2}
R_{ij} - \frac{1}{2}g_{ij}R =  8\pi G \; T_{ij} + \Lambda \; g_{ij},
\end{equation}
where the symbols have their usual meaning. \\

For a perfect fluid, the stress-energy-momentum tensor $T_{ij}$ is given by
\begin{equation}
\label{eq3}
T_{ij} = (\rho + p)u_{i}u_{j} - p\;g_{ij} ,
\end{equation}
where $\rho$ is the matter density, p is the thermodynamics pressure and $u^{i}$ is the fluid four-velocity
vector of the fluid satisfying the condition
\begin{equation}
\label{eq4}
u^{i}u_{i} =  1.
\end{equation}
In the field Eqs. (\ref{eq2}), $\Lambda$ accounts for vacuum energy with its energy density $\rho_{v}$ and
pressure $p_{v}$ satisfying the equation of state
\begin{equation}
\label{eq5}
p_{v} = -\rho_{v} = -\frac{\Lambda}{8\pi G}
\end{equation}
The critical density and the density parameters for matter and cosmological constant are, respectively, defined as
\begin{equation}
\label{eq6}
\rho_{c} = \frac{3H^{2}}{8\pi G},
\end{equation}
\begin{equation}
\label{eq7}
\Omega_{M} = \frac{\rho}{\rho_{c}} = \frac{8\pi G \rho}{3H^{2}},
\end{equation}
\begin{equation}
\label{eq8}
\Omega_{\Lambda} = \frac{\rho_{v}}{\rho_{c}} = \frac{\Lambda}{3H^{2}}.
\end{equation}
We observe that the density parameters $\Omega_{M}$ and $\Omega_{\Lambda}$ are singular when H = 0. \\ 

In a comoving system of coordinates, the field Eqs. (\ref{eq2}) for the metric (\ref{eq1}) with (\ref{eq3}) read as
\begin{equation}
\label{eq9}
\frac{\ddot{A}}{A} + \frac{\ddot{B}}{B} + \frac{\dot{A}\dot{B}}{AB} = - 8\pi G p + \Lambda,
\end{equation}
\begin{equation}
\label{eq10}
\frac{\ddot{A}}{A} + \frac{\ddot{C}}{C} + \frac{\dot{A}\dot{C}}{AC} = - 8\pi G p + \Lambda,
\end{equation}
\begin{equation}
\label{eq11}
\frac{\ddot{B}}{B} + \frac{\ddot{C}}{C} + \frac{\dot{B}\dot{C}}{BC} = - 8\pi G p + \Lambda,
\end{equation}
\begin{equation}
\label{eq12}
\frac{\dot{A}\dot{B}}{AB} + \frac{\dot{B}\dot{C}}{BC} + \frac{\dot{C}\dot{A}}{CA} = 8\pi G \rho + \Lambda.
\end{equation}
The covariant divergence of Eq. (\ref{eq2}) yields
\begin{equation}
\label{eq13}
\dot{\rho} + 3(\rho + p)H + \rho \frac{\dot G}{G} + \frac{\dot{\Lambda}}{8\pi G} = 0.
\end{equation}
Spatial volume for the model given by Eq. (\ref{eq1}) reads as
\begin{equation}
\label{eq14}
V = ABC
\end{equation}
We define average scale factor a of anisotropic model as
\begin{equation}
\label{eq15}
a = (ABC)^{\frac{1}{3}} = V^{\frac{1}{3}}.
\end{equation}
So that generalized mean Hubble parameter $H$ is given by
\begin{equation}
\label{eq16}
H = \frac{1}{3}(H_{x} + H_{y} + H_{z}) ,
\end{equation}
where $H_{x} = \frac{\dot A}{A}, H_{y} = \frac{\dot B}{B}, H_{z} = \frac{\dot C}{C}$
are the directional Hubble parameters in direction of x, y and z respectively and a dot denotes
differentiation with respect to cosmic time t. \\

From Eqs. (\ref{eq15}) and (\ref{eq16}), we obtain an important relation
\begin{equation}
\label{eq17}
H = \frac{\dot a}{a} = \frac{1}{3} \left(\frac{\dot A}{A} + \frac{\dot B}{B} + \frac{\dot C}{C}\right) .
\end{equation}
Expressions for the dynamical scalars such as the expansion scalar $(\theta)$, anisotropy parameter
$(A_{m})$ and the shear scalar $(\sigma)$ are defined as usual:
\begin{equation}
\label{eq18}
\theta = u^{i}_{;i} = \left(\frac{\dot A}{A} + \frac{\dot B}{B} + \frac{\dot C}{C}\right) ,
\end{equation}
\begin{equation}
\label{eq19}
A_{m} = \frac{1}{3}\sum_{i=1}^{3}\left(\frac{H_{i} - H}{H}\right)^{2},
\end{equation}
\begin{equation}
\label{eq20}
\sigma^{2} = \frac{1}{2}\sigma_{ij}\sigma^{ij} = \frac{1}{2}\left[\left(\frac{\dot A}{A}\right)^{2}
+ \left(\frac{\dot B}{B}\right)^{2} + \left(\frac{\dot C}{C}\right)^{2}\right] - \frac{\theta^{2}}{6} .
\end{equation}
We define deceleration parameter (DP) q as
\begin{equation}
\label{eq21}
q = -\frac{a \ddot{a}}{\dot{a}^{2}} = -\left(\frac{\dot{H} + H^{2}}{H^{2}}\right).
\end{equation}

\section{Solution of field equations}
The field Eqs. (\ref{eq9})$-$(\ref{eq12}) are a system of four equations with seven unknown parameters 
$A$, $B$, $C$, $G$, $p$, $\rho$ and $\Lambda$. Hence, three additional constraints relating these parameters 
are required to obtain explicit solution of the system.\\ 

So first, we assume a power-law form of the gravitational constant ($G$) with scale factor $a$ as proposed by 
Chawla et al. \cite{ref40}  
\begin{equation}
\label{eq22}
G \propto a^{m},
\end{equation}
where m is a constant. For the sake of mathematical simplicity, Eq. (\ref{eq22}) may be written as
\begin{equation}
\label{eq23}
G = G_{0} a^{m},
\end{equation}
where $G_{0}$ is a positive constant.\\ 

Secondly, we assume equation of state for perfect fluid as
\begin{equation}
\label{eq24}
p = \gamma \rho,  
\end{equation}
where $\gamma$ ($0 \leq \gamma \leq 1$) is constant. \\

Following the technique [44, 61$-$63], we get three equations 
from field Eqs. (\ref{eq9})$-$(\ref{eq11})
\begin{equation}
\label{eq25}
\frac{A}{B} = d_{1} \exp \left(k_{1} \int {a^{-3}dt} \right),
\end{equation}
\begin{equation}
\label{eq26}
\frac{B}{C} = d_{2} \exp \left(k_{2} \int {a^{-3}dt} \right),
\end{equation}
\begin{equation}
\label{eq27}
\frac{C}{A} = d_{3} \exp \left(k_{3} \int {a^{-3}dt} \right),
\end{equation}
where $d_{1}, d_{2}, d_{3}$ and $k_{1}, k_{2}, k_{3}$ are constants of integration.
Finally, using $a = (ABC)^{\frac{1}{3}}$, we write the metric functions from Eqs. (\ref{eq25})$-$(\ref{eq27})
in explicit form as
\begin{equation}
\label{eq28}
A(t) = l_{1} a \exp \left(m_{1} \int {a^{-3}dt} \right),
\end{equation}
\begin{equation}
\label{eq29}
B(t) = l_{2} a \exp \left(m_{2} \int {a^{-3}dt} \right),
\end{equation}
\begin{equation}
\label{eq30}
C(t) = l_{3} a \exp \left(m_{3} \int {a^{-3}dt} \right),
\end{equation}
where constants $m_{1}, m_{2}, m_{3}$ and $l_{1}, l_{2}, l_{3}$ satisfy the fallowing two relations:
\begin{equation}
\label{eq31}
m_{1} + m_{2} + m_{3} = 0, \; \; \; l_{1}l_{2}l_{3} = 1.
\end{equation}
in the particular case  
\begin{equation}
\label{eq32}
l_{1} = \sqrt[3]{d_{1}d_{2}}, \indent l_{2} = \sqrt[3]{d_{1}^{-1}d_{3}}, \indent l_{3} =
\sqrt[3]{(d_{2}d_{3})^{-1}},
\end{equation}
and
\begin{equation}
\label{eq33}
m_{1} = \frac{k_{1} + k_{2}}{3}, \indent  m_{2} = \frac{k_{3} - k_{1}}{3}, \indent  m_{3} =
\frac{-(k_{2} + k_{3})}{3}.
\end{equation}
Now, the metric functions can be determined as functions of cosmic time t if the average scale factor is known. Hence, we 
consider the {\it ansatz} for the scale factor, where increase in term of time evolution is
\begin{equation}
\label{eq34}
a = (t^{k} e^{t})^\frac{1}{n}
\end{equation}
In our earlier studies it is found that {\it ansatz} generalizes [52$-$55]. The choice of scale factor attracts a time-dependent 
deceleration parameter (see Eq. (\ref{eq47})) which brings that dark energy era, the  solution gives inflation and radiation/matter 
dominance era with subsequent transition from deceleration to acceleration. Now for a Universe which was decelerating in past 
and accelerating at the present time, the deceleration parameter must show signature flipping \cite{ref56}. This theme motivates to 
choose such scale factor (\ref{eq34}) that yields a time dependent deceleration parameter given by Eq. (\ref{eq47}). \\

Using Eq. (\ref{eq34}) in Eqs. (\ref{eq28})$-$(\ref{eq30}), we obtain
\begin{equation}
\label{eq35}
A(t) = l_{1} (t^{k} e^{t})^{\frac{1}{n}} \exp{\left [m_{1} F(t)\right]},
\end{equation}
\begin{equation}
\label{eq36}
B(t) = l_{2} (t^{k} e^{t})^{\frac{1}{n}} \exp{\left [m_{2} F(t)\right]},
\end{equation}
\begin{equation}
\label{eq37}
C(t) = l_{3} (t^{k} e^{t})^{\frac{1}{n}} \exp{\left [m_{3} F(t)\right]},
\end{equation}
where 
\begin{equation}
\label{eq38}
F(t) = \int{(t^{k} e^{t})^{-\frac{3}{n}dt}} = \sum_{i=1}^{\infty} \frac{(-3)^{i-1} t^{i-\frac{3k}{n}}}{n^{i-2}(ni-3k)(i-1)!}.
\end{equation}
From Eqs. (\ref{eq35})$-$(\ref{eq37}), we obtain
\[
\frac{\dot{A}}{A} = \frac{1}{n} \left(\frac{k}{t}+1 \right) + m_{1}(t^{k}e^{t})^{-\frac{3}{n}},
\]
\[
\frac{\dot{B}}{B} = \frac{1}{n} \left(\frac{k}{t}+1 \right) + m_{2}(t^{k}e^{t})^{-\frac{3}{n}},
\]
\begin{equation}
\label{eq39}
\frac{\dot{C}}{C}= \frac{1}{n} \left(\frac{k}{t}+1 \right) + m_{3}(t^{k}e^{t})^{-\frac{3}{n}},
\end{equation}
and \\
\[
\frac{\ddot{A}}{A} = \frac{1}{n^{2}}\left(\frac{k}{t}+1 \right)^{2} - \frac{k}{nt^{2}} + m_{1}^{2} (t^{k}e^{t})^{-\frac{6}{n}} 
-\frac{m_{1}}{n}(t^{k}e^{t})^{-\frac{3}{n}}\left(\frac{k}{t} + 1 \right), 
\]
\[
\frac{\ddot{B}}{B} =\frac{1}{n^{2}}\left(\frac{k}{t}+1 \right)^{2} - \frac{k}{nt^{2}} + m_{2}^{2} (t^{k}e^{t})^{-\frac{6}{n}} 
-\frac{m_{2}}{n}(t^{k}e^{t})^{-\frac{3}{n}}\left(\frac{k}{t} + 1 \right), 
\]
\begin{equation}
\label{eq40}
\frac{\ddot{C}}{C} = \frac{1}{n^{2}}\left(\frac{k}{t}+1 \right)^{2} - \frac{k}{nt^{2}} + m_{3}^{2} (t^{k}e^{t})^{-\frac{6}{n}} 
-\frac{m_{3}}{n}(t^{k}e^{t})^{-\frac{3}{n}}\left(\frac{k}{t} + 1 \right). 
\end{equation}
Hence the geometry of the universe (\ref{eq1}) is reduced to
\[
ds^{2} = -dt^{2} + (t^{k} e^{t})^{\frac{2}{n}} \Biggl[l_{1}^{2}\exp{\{2m_{1}F(t)\}}dx^{2} + l_{2}^{2} \exp{\{2m_{2}F(t)\}}dy^{2}
\]
\begin{equation}
\label{eq41}
 + l_{3}^{2} \exp{\{2m_{3}F(t)\}}dz^{2}\Biggr].
\end{equation}

\section{Results and discussion}
Expressions for physical parameters such as spatial volume ($V$), mean Hubble's parameter ($H$), expansion scalar ($\theta$), 
shear scalar ($\sigma$) and anisotropy parameter ($A_{m}$) for model (\ref{eq41}) are given by
\begin{equation}
\label{eq42}
V = (t^{k}e^{t})^{\frac{3}{n}},
\end{equation}
\begin{equation}
\label{eq43}
H = \frac{1}{n}\left(\frac{k}{t} + 1\right),
\end{equation}
\begin{equation}
\label{eq44}
\theta = \frac{3}{n}\left(\frac{k}{t} + 1\right),
\end{equation}
\begin{equation}
\label{eq45}
A_{m} = \frac{\beta _{1}}{3n^{2}} \left(\frac{k}{t} + 1 \right)^{-2}(t^{k}e^{t})^{-\frac{6}{n}},
\end{equation}
\begin{equation}
\label{eq46}
\sigma^{2} = \frac{\beta_{1}}{2}(t^{k}e^{t})^{-\frac{6}{n}},
\end{equation}
where \\
$\beta_{1} = m_{1}^{2} + m_{2}^{2} + m_{3}^{2}$. \\

\begin{figure}[ht]
\begin{minipage}[b]{0.4\linewidth}
\centering
\includegraphics[width=\textwidth]{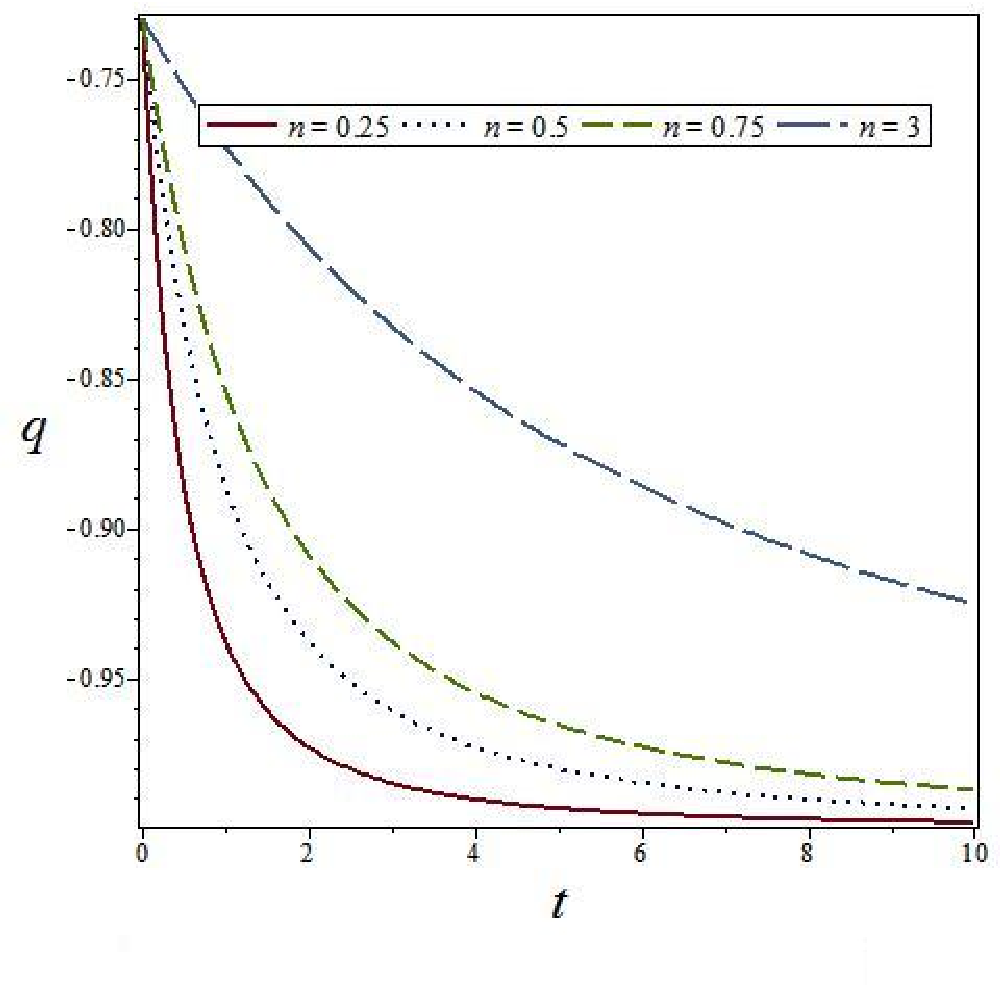}
\caption{Plots of deceleration parameter $q$ versus time $t$}
\label{fig:figure1}
\end{minipage}
\hspace{0.5cm}
\begin{minipage}[b]{0.4\linewidth}
\centering
\includegraphics[width=\textwidth]{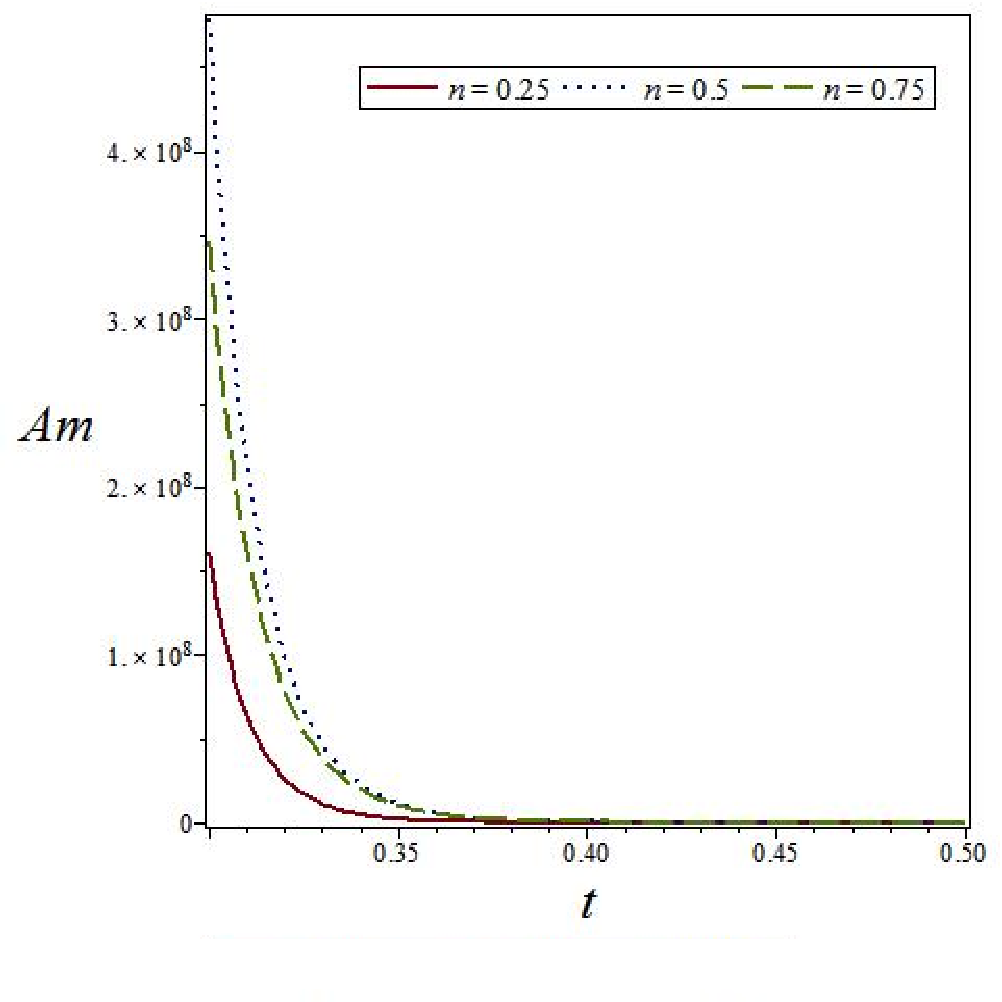}
\caption{Plots of anisotropic parameter $A_{m}$ versus time $t$ for $m_{1} = 0.25$, $m_{2} = 0.75$, $m_{3} = -1$}
\label{fig:figure2}
\end{minipage}
\end{figure}

From Eq. (\ref{eq21}), the deceleration parameter is computed as
\begin{equation}
\label{eq47}
q = -1 + \frac{n k}{(k+t)^{2}}.
\end{equation}
From Eq. (\ref{eq47}), we observe that $q > 0$ for $t < \sqrt{nk} - k$ and $q < 0$ for $t > \sqrt{nk} - k$. It is 
observed that for $n \geq 3 ~ \& ~ k = 1$, our model is evolving from decelerating phase to accelerating phase. Also, 
recent observations of SNe Ia expose that the present universe is accelerating and the value of DP lies on some place 
in the range $-1 < q < 0$. It follows that in our derived model, one can choose the value of DP consistent with the 
observation. Fig. $1$ depicts the variation of deceleration parameter ($q$) with cosmic time, giving the behaviour 
of $q$ as in accelerating phase at present epoch for different values of $(n, k)$ which is consistent with recent observations 
of Type Ia supernovae \cite{ref1,ref2}. \\

From above Eq. (\ref{eq47}), present value of deceleration parameter can be estimated as
\begin{equation}
\label{eq48} q_{0} = -1 + \frac{k}{nH_{0}^{2}t_{0}^{2}},
\end{equation}
where where $H_{0}$ is present value of Hubble's parameter and $t_{0}$ is the age of universe at present epoch. Recent 
observations show that the deceleration parameter of the universe is in the range $ - 1 \leq q \leq 0$ i.e $q_{0} \approx -0.77$.
For $n = 0.27k$, we obtain $q_{0} = -0.73$ which is similar to the observed value of DP at present epoch \cite{ref57}. Therefore, 
we restrict the values of $n$ and $k$ such that the condition $n = 0.27k$ is satisfied for graphical representations of the 
physical parameters. As a representative case, we have considered four values of $(n,k)$ as (0.25, 0.9259259260), (0.50, 0.1.851851852), 
(0.75, 2.77777778) and (3, 11.11111111) respectively for graphic presentation of Figs. 1$-$8.\\

From Eqs. (\ref{eq42}) and (\ref{eq44}) we observe that the spatial volume is zero at $t = 0$ and the expansion scalar 
is infinite, which show that the universe starts evolving with zero volume at $t = 0$ which is big bang scenario. From 
Eqs. (\ref{eq35})$-$(\ref{eq37}), we observe that the spatial scale factors are zero at the initial epoch $t = 0$ and hence 
the model has a point type singularity \cite{ref58}. We observe that proper volume increases with time. \\
 
From Eq. (\ref{eq45}), we observe that at late time when $t \to \infty$, $A_{m} \to 0$. Thus, our model has transition 
from initial anisotropy to isotropy at present epoch which is in good harmony with current observations. Fig. $2$ depicts 
the variation of anisotropic parameter ($A_{m}$) with cosmic time $t$. From the figure, we observe that $A_{m}$ decreases 
with time and tends to zero as $t \to \infty$. Thus, the observed isotropy of the universe can be achieved in our model at 
present epoch. \\

It is important to note here that $\lim_{t \to 0}\left(\frac{\rho}{\theta^{2}}\right)$ spread out to be constant. Therefore, the 
model of the universe goes up homogeneity and matter is dynamically negligible near the origin. This is in good agreement with 
the result already given by Collins \cite{ref59}. \\ 

Using Eq. (\ref{eq34}) into (\ref{eq23}), the gravitational constant is obtained as
\begin{equation}
\label{eq49}
G = G_{0}(t^{k}e^{t})^{\frac{m}{n}}.
\end{equation}
\begin{figure}[ht]
\begin{minipage}[b]{0.4\linewidth}
\centering
\includegraphics[width=\textwidth]{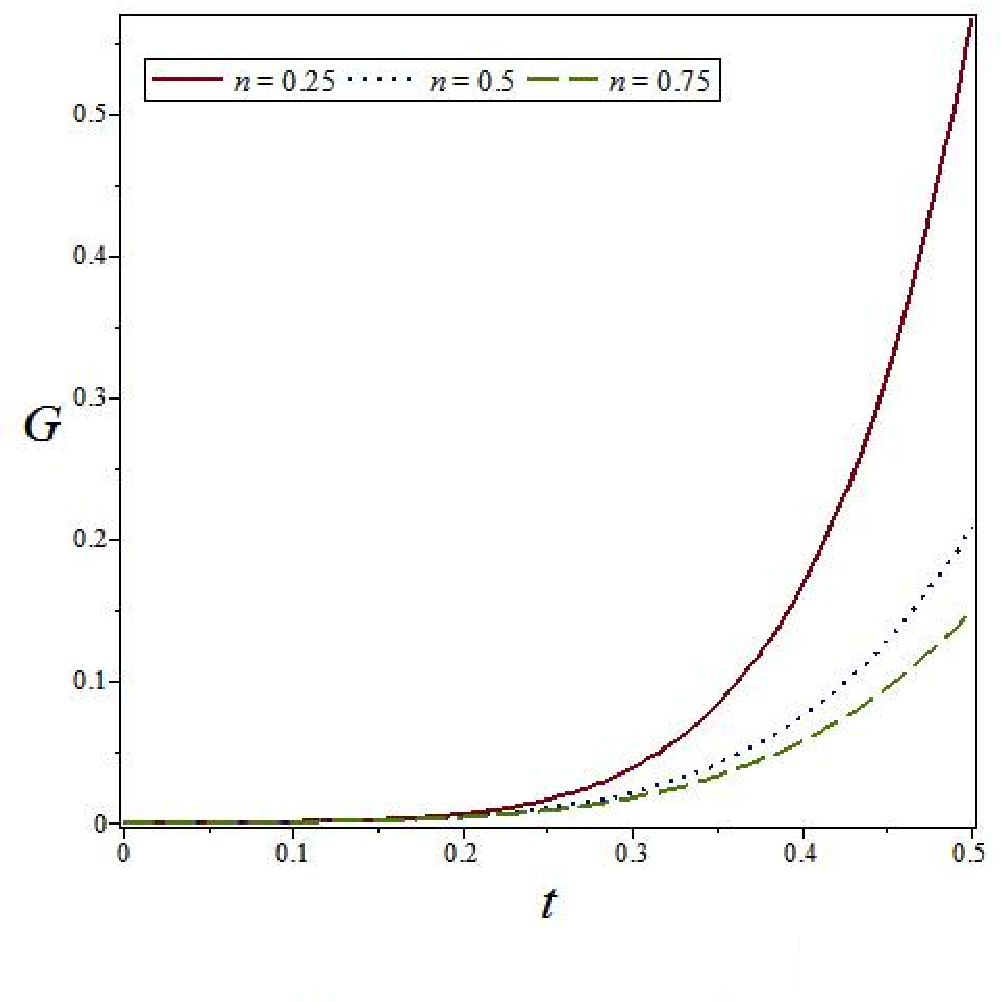}
\caption{Plots of gravitational constant $G$ versus time $t$ for $G_{0} = m = 1$}
\label{fig:figure3}
\end{minipage}
\hspace{0.5cm}
\begin{minipage}[b]{0.4\linewidth}
\centering
\includegraphics[width=\textwidth]{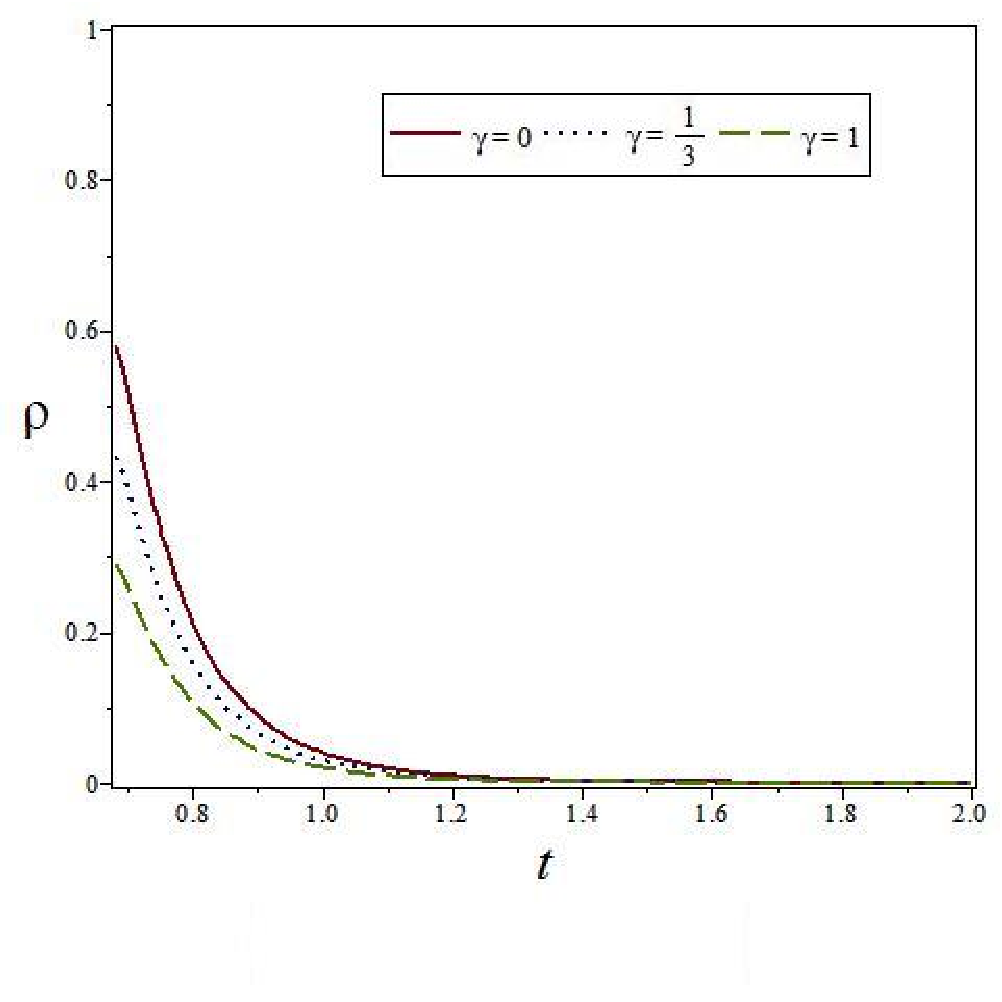}
\caption{Plots of energy density $\rho$ versus time $t$ for $m = 1$, $m_{1} = 0.25$, $m_{2} = 0.75$, $m_{3} = -1$, $n = 0.5$}
\label{fig:figure4}
\end{minipage}
\end{figure}


\begin{figure}[ht]
\begin{minipage}[b]{0.4\linewidth}
\centering
\includegraphics[width=\textwidth]{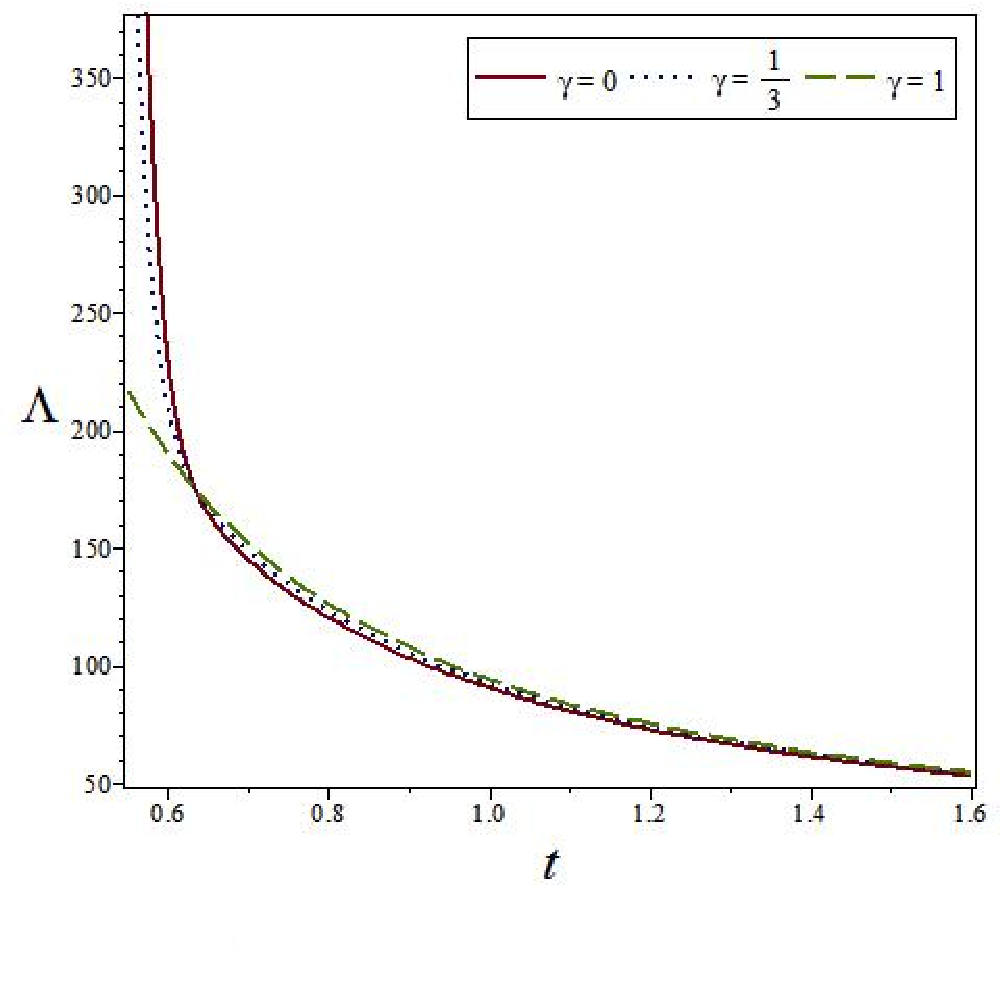}
\caption{Plots of cosmological constant $\Lambda$ versus time $t$ for $m_{1} = 0.25$, $m_{2} = 0.75$, $m_{3} = -1$, $n = 0.5$}
\label{fig:figure5}
\end{minipage}
\hspace{0.5cm}
\begin{minipage}[b]{0.4\linewidth}
\centering
\includegraphics[width=\textwidth]{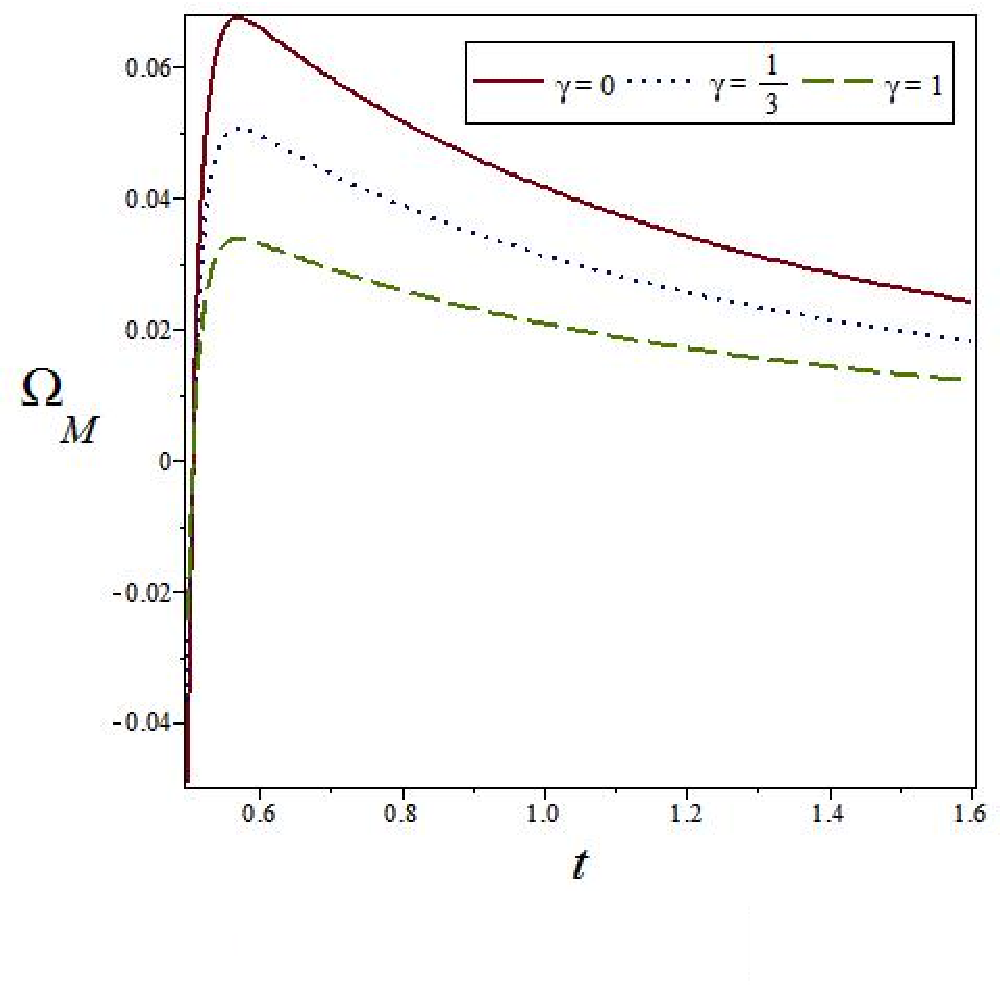}
\caption{Plots of density parameter $\Omega_{M}$ versus time $t$ for $m_{1} = 0.25$, $m_{2} = 0.75$, $m_{3} = -1$, $n = 0.5$}
\label{fig:figure6}
\end{minipage}
\end{figure}
\begin{figure}[ht]
\begin{minipage}[b]{0.4\linewidth}
\centering
\includegraphics[width=\textwidth]{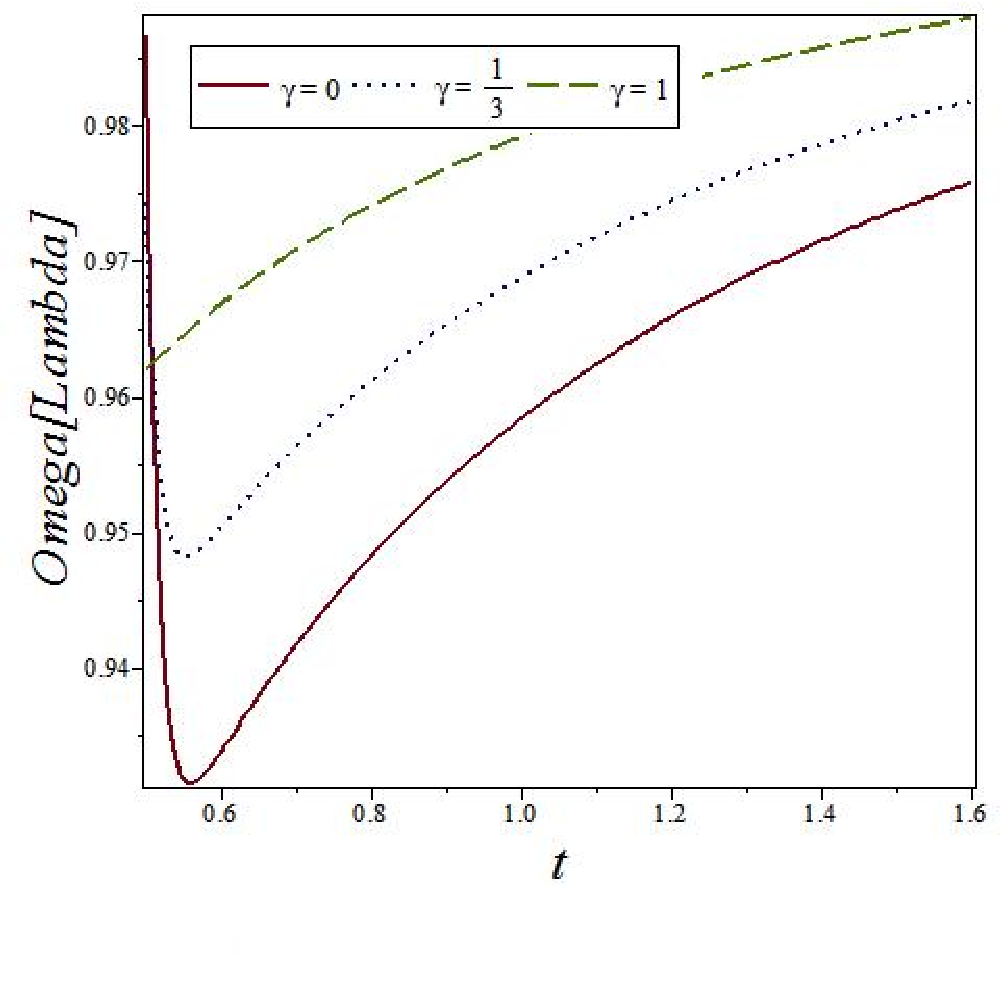}
\caption{Plots of density parameter $\Omega_{\Lambda}$ versus time $t$ for $m_{1} = 0.25$, $m_{2} = 0.75$, $m_{3} = -1$, $n = 0.5$}
\label{fig:figure7}
\end{minipage}
\hspace{0.5cm}
\begin{minipage}[b]{0.4\linewidth}
\centering
\includegraphics[width=\textwidth]{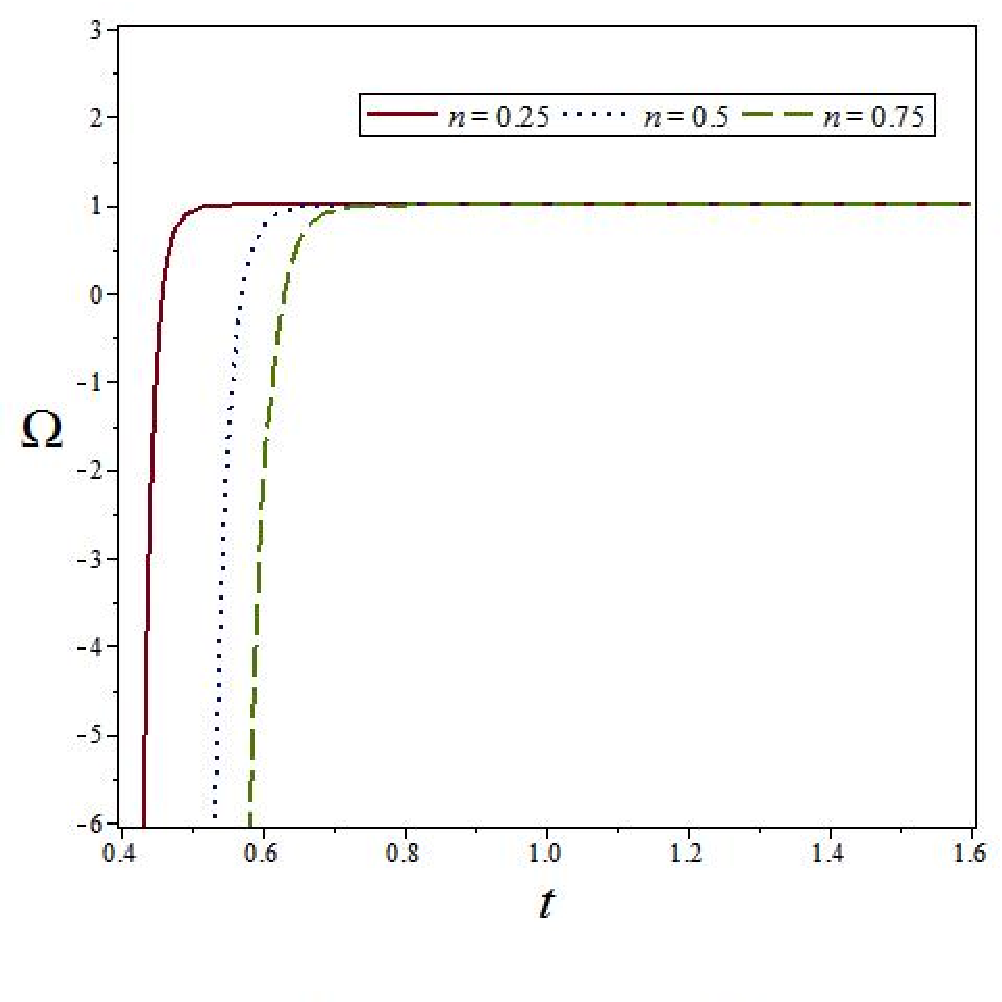}
\caption{Plots of total density parameter $\Omega$ versus time $t$ for $m_{1} = 0.25$, $m_{2} = 0.75$, $m_{3} = -1$}
\label{fig:figure8}
\end{minipage}
\end{figure}
From Eq. (\ref{eq49}), we observe that $G \to 0$ as $t \to 0$ whereas for $t \to \infty$, $G \to \infty$ which shows that $G$ 
is an increasing function of time. This nature of variation of $G$ with cosmic time is shown in Fig. $3$ for three values 
of $n = 0.25, 0.50$ and $0.75$. Abdel-Rahaman \cite{ref60} has investigated cosmological models in which $G$ increases in a 
constantly expanding universe. Some authors \cite{ref61,ref62} found in their study that G is decreasing function of time whereas 
an increasing $G$ has also been reported [36, 37, 63]. This alteration in $G$ leaves the form of Einstein's equations formally 
intact by allowing the variation of $G$ to be accompanied by a change in $\Lambda$ and enables to solve the cosmological constant 
problem, inflationary scenario etc. \cite{ref64}. \\

Using Eqs. (\ref{eq24}), (\ref{eq39}) and (\ref{eq40}) and solving the field Eqs. (\ref{eq9})$-$(\ref{eq12}), we get the 
expressions for energy density, pressure and cosmological constant for universe (\ref{eq41}) as 
\begin{equation}
\label{eq50}
\rho = \frac{1}{8\pi G_{0} (1 + \gamma)} \left[\frac{2k}{n t^{2}}(t^{k}e^{t})^{-\frac{m}{n}} - 
\beta_{1}(t^{k}e^{t})^{-\frac{(m+6)}{n}}\right],
\end{equation}
\begin{equation}
\label{eq51}
p = \frac{\gamma}{8\pi G_{0} (1 + \gamma)} \left[\frac{2k}{n t^{2}}(t^{k}e^{t})^{-\frac{m}{n}} - 
\beta_{1}(t^{k}e^{t})^{-\frac{(m+6)}{n}}\right],
\end{equation}
\begin{equation}
\label{eq52}
\Lambda = \frac{3}{n^{2}}\left(\frac{k}{t} + 1 \right)^{2} + \frac{1}{(1 + \gamma)}\left[\beta_{2} (t^{k}e^{t})^{-\frac{6}{n}} - 
\frac{2k}{nt^{2}}\right],
\end{equation}
where \\
$\beta_{2} = m_{1}^{2} + m_{2}^{2}+m_{1}m_{2} + \gamma(m_{1}m_{2}+m_{2}m_{3}+m_{3}m_{1})$. \\ 

We find that the above solutions satisfy Eq. (\ref{eq13}) identically and hence represent exact solution of Einstein's field 
equations (\ref{eq9})$-$(\ref{eq12}). \\

From above relations (\ref{eq50})$-$(\ref{eq52}), we can obtain the expressions of energy density, pressure and cosmological 
constant for four types of models: (i) Empty model for $\gamma = 0$, (ii) radiating dominated model for $\gamma = \frac{1}{3}$, 
(iii) degenerate vacuum or false vacuum or $\rho$ vacuum model \cite{ref65} for $\gamma = -1$ and (iv) Zeldovich fluid or stiff 
fluid model \cite{ref66,ref67} for $\gamma = 1$. \\

From Eq. (\ref{eq50}), it is observed that the energy density $\rho$ is a decreasing function of time and 
$\rho > 0$ under condition $t^{3k -1}e^{\frac{6t}{n}} > \frac{n\beta_{1}}{2k}$. The energy density 
versus time is shown in Fig. $4$ for $\gamma = 0, \frac{1}{3}$ and $1$. It is evident that the energy density 
remains positive in all the three types of models under appropriate condition. However, it decreases more 
sharply with the cosmic time in Zeldovich universe, compared to radiating dominated and empty fluid universes. \\

Fig. $5$ shows the plot of the cosmological term $\Lambda$ versus cosmic time for $\gamma = 0, \frac{1}{3}$ and $1$. We 
observe that $\Lambda$ is decreasing function of time $t$ and it approaches a small positive value at late time 
in all the three types of models. We also observe that $\Lambda$ decreases more aggressively with the cosmic time 
in empty universe than radiating dominated and stiff fluid universes. From recent cosmological observations 
\cite{ref1,ref3,ref5}, it is well established that cosmological constant $\Lambda$ is small and positive at 
present epoch. A positive $\Lambda$ increases the expansion speed with time while a negative $\Lambda$ tends 
to decelerate it. Thus, the nature of $\Lambda$ in our derived models are supported by recent observations. \\

The vacuum energy density ($\rho_{\nu}$), critical density $(\rho_{c})$ and the density parameters $(\Omega_{M}, \Omega_{\Lambda})$ 
for model (\ref{eq41}) read as
\[
\rho_{\nu}= \frac{1}{8\pi G_{0}} \Biggl[\frac{3}{n^{2}}\left(\frac{k}{t} + 1 \right)^{2}(t^{k}e^{t})^{-\frac{m}{n}} + 
\frac{\beta_{2}}{(1 + \gamma)} (t^{k}e^{t})^{-\frac{(m+6)}{n}}
\]
\begin{equation}
\label{eq53}
 - \frac{2k}{n(1+\gamma)} t^{-\frac{km}{n} + 2} e^{-\frac{mt}{n}}\Biggr],
\end{equation}
\begin{equation}
\label{eq54}
\rho_{c} = \frac{3}{8\pi G_{0}n^{2}} \left(\frac{k}{t} + 1 \right)^{2}(t^{k}e^{t})^{-\frac{m}{n}},
\end{equation}
\begin{equation}
\label{eq55}
\Omega_{M} = \frac{n^{2}\left[\frac{2k}{nt^{2}} - \beta_{1}(t^{k}e^{t})^{-\frac{6}{n}}\right]}{3(1+\gamma)\left(\frac{k}{t} 
+ 1 \right)^{2}},
\end{equation}
\begin{equation}
\label{eq56}
\Omega_{\Lambda} = 1 + \frac{n^{2}\left[\beta_{2}(t^{k}e^{t})^{-\frac{6}{n}} - \frac{2k}{nt^{2}}\right]}{3 (1+ \gamma)(\frac{k}{t} 
+ 1)^{2}}.
\end{equation}
Adding Eqs. (\ref{eq55}) and (\ref{eq56}), we get
\begin{equation}
\label{eq57}
\Omega = \Omega_{M} + \Omega_{\Lambda} = 1 + \frac{\beta n^{2}(t^{k}e^{t})^{-\frac{6}{n}}}{3\left(\frac{k}{t} + 1\right)^{2}},
\end{equation}
where $\beta =  m_{1}m_{2} + m_{2}m_{3} +m_{3}m_{1}$. For $\beta = 0$, we have $\Omega = 1$. We also observe from 
Eq. (\ref{eq57}) that $\Omega $ approaches to one for sufficiently large time independent to $\beta$. Figs. $6$ and $7$ 
plot the variation of density parameters for matter ($\Omega_{M}$) and cosmological constant ($\Omega_{\Lambda}$) versus $t$ 
respectively. From these figures it is clear that the universe is dominated by matter in early stage of evolution whereas 
it is dominated by dark energy (cosmological constant $\Lambda$) at present epoch. Fig. $8$ shows a variation 
of total energy parameter ($\Omega$) versus cosmic time $t$. From the Fig. $8$, we observe that $\Omega \to 1$ at late time 
for arbitrary value of $\beta$. This is in good agreement with the observational results \cite{ref68}. \\
\begin{figure}[ht]
\begin{minipage}[b]{0.4\linewidth}
\centering
\includegraphics[width=\textwidth]{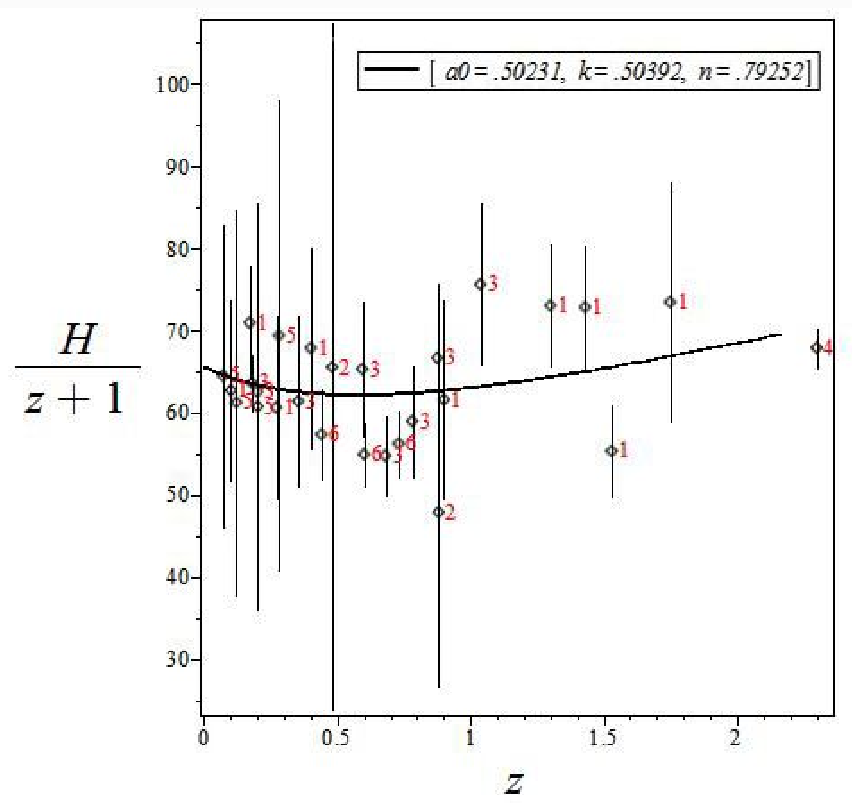}
\caption{H(z)/(1+z) data (32 points) and model prediction as a function of redshift.}
\label{fig:figure9}
\end{minipage}
\hspace{0.5cm}
\begin{minipage}[b]{0.4\linewidth}
\centering
\includegraphics[width=\textwidth]{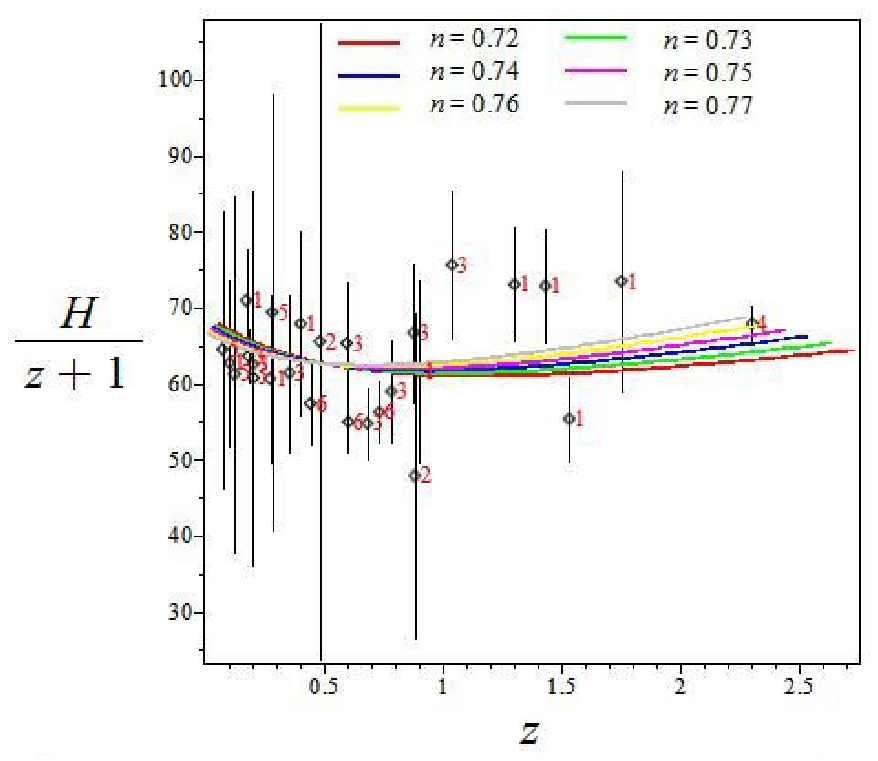}
\caption{H(z)/(1+z) data (32 points) and model prediction as a function of redshift for different $n$ with fixed $k = 0.5$ and $a_0 = 0.5$.}
\label{fig:figure10}
\end{minipage}
\end{figure}

\begin{figure}[ht]
\begin{minipage}[b]{0.4\linewidth}
\centering
\includegraphics[width=\textwidth]{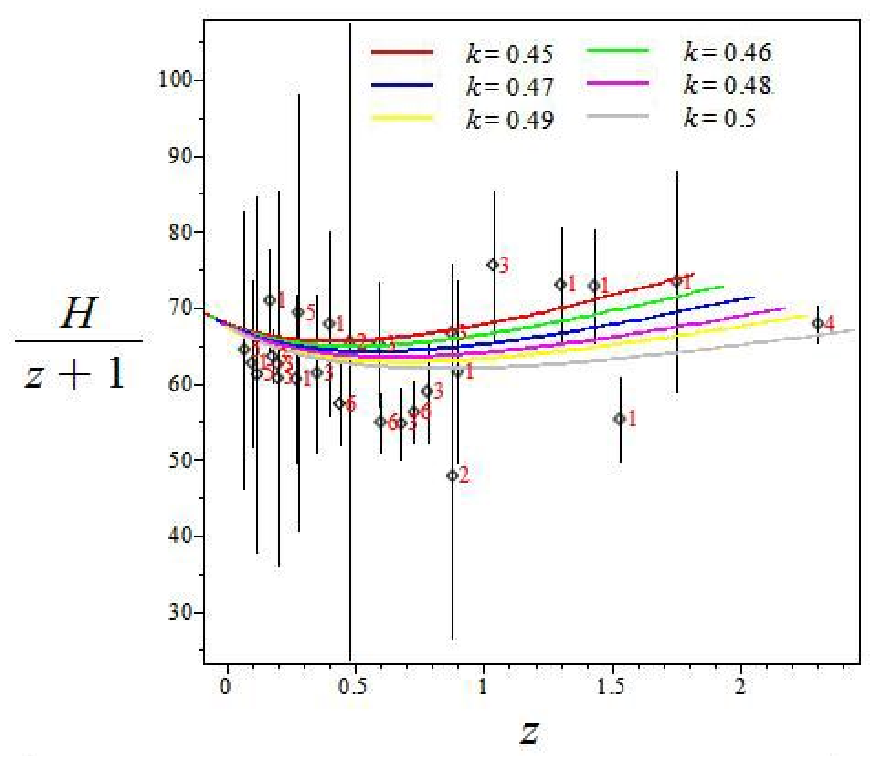}
\caption{H(z)/(1+z) data (32 points) and model prediction as a function of redshift for different $k$ with fixed $n = 0.75$ and $a_0 = 0.5$. }
\label{fig:figure11}
\end{minipage}
\hspace{0.5cm}
\begin{minipage}[b]{0.4\linewidth}
\centering
\includegraphics[width=\textwidth]{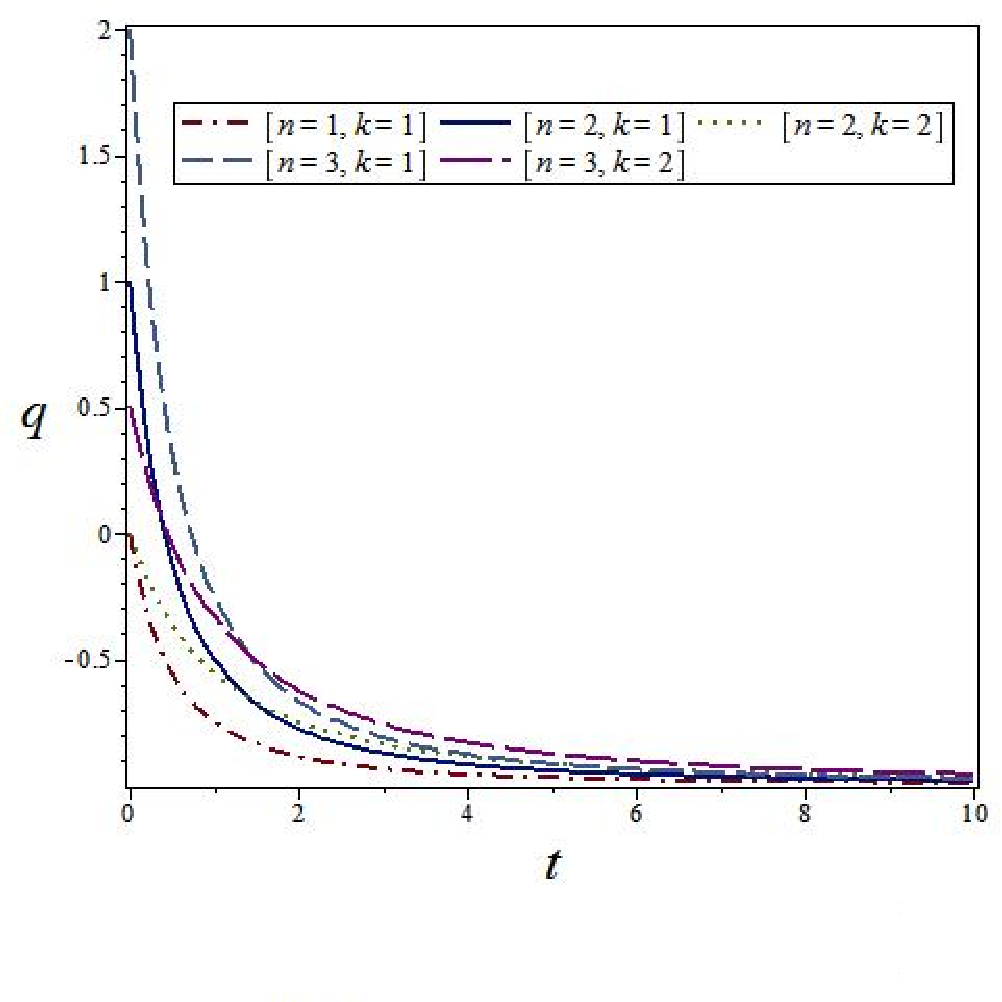}
\caption{ Plots of deceleration parameter $q$ vs time $t$ for transit universe for different sets of $k$ and $n$.}
\label{fig:figure12}
\end{minipage}
\end{figure}

{\bf{The updated Hubble parameter measurements:}} \\

Recently Farooq and Ratra \cite{ref69}, following Farooq et al. \cite{ref70}, have compiled a list of $28$ independent 
measurements of the Hubble parameter between redshift $0.07 \leq z \leq 2.3$ and used this to place constraints on model 
parameter of constant and time-evolving dark energy cosmologies. This is the extension of the analysis of Busca et al. 
\cite{ref71} to a larger independent set and determined the cosmological deceleration-acceleration transition redshift 
$z_{da} = 0.74 \pm 0.05$. These $H(z)$ data are well-described by all best-fit models, and provide tight constraints on 
the model parameter. In the following we plotted $H(z)/(1+z)$ as a function of $z$, and also plotted the data (Table 1) of 
Farooq and Ratra \cite{ref69} in Figs. [9$-$11]. This was determined from the $6$ best-fit transition redshifts measured 
in three different cosmological models, $\Lambda$CDM, XCDM, and $\phi$CDM, for two different Hubble constant priors.\\

The expansion rate, i.e. the Hubble parameter, is defined as: \\
\begin{equation}
\label{eq58}
H(z) = \frac{\dot a}{a} = - \frac{1}{1 + z}\frac{dz}{dt}, 
\end{equation}
where the redshift $z$ of the chronometers is known with a high accuracy based on the spectra of galaxies and the differential 
measurement of time ($dt$) at a given redshift interval ($dz$) automatically provides a direct measurement on $H(z)$. \\

From $a (t) = (t^k e^t)^{1/n}$ we have
\begin{equation}
\label{eq59}
H = \frac{1}{n}\left(\frac{k}{t} + 1 \right),
\end{equation}
and
\begin{equation}
\label{eq60}
 1+ z = \frac{a (t_0)}{a(t)} = \frac{a_0}{(t^k e^t)^{1/n}}.
\end{equation}
Suppose $H(z)$ is theoretical function. $H_i$ is measured at point $z_i$ (experimental data taken from the paper of 
Farooq and Ratra \cite{ref69}) differs from $H(z_i)$ by random values having Gaussian distribution with zero
mathematical expectation and standard deviation $\sigma_i$. The expression
\begin{equation}
\label{eq61}
 \xi_i = \frac{H(z_i) - H_i}{\sigma_{H_i}} 
\end{equation}
possesses Gaussian distribution with zero mathematical expectation and unit standard
deviation. The quantities
\begin{equation}
\label{eq62}
\chi^2 = \sum_i \xi_i^2
\end{equation}
have known $\chi^2$ distribution and maximum likelihood is attained by minimizing $\chi^2$ by parameters $a_0, k, n$. \\

The minimum was reached at $a_0 = 0.502$, $k = 0.504$ and $n = 0.793$ with the magnitude

\begin{equation}
\label{eq63}
\min_{a_0, k, n} \sqrt{\frac{1}{N} \sum_{i =1}^{N}\Biggl(\frac{H(z_i) - H_i}{\sigma_{H_i}}\Biggr)^2} = 0.8
\end{equation}
The $H(z)$ data require accelerated cosmological expansion at the current epoch, and are consistent with the 
decelerated cosmological expansion at earlier times predicted and required in standard dark energy models. \\

Fig. $9$ shows the H(z)/(1+z) data (32 points) and model prediction as a function of redshift. Fig. $10$ represents  H(z)/(1+z) 
data ($32$ points) and model prediction as a function of redshift for different $n$ with fixed $k = 0.5$ and $a_0 = 0.5$ whereas 
Fig. $11$ indicates H(z)/(1+z) data (32 points) and model prediction as a function of redshift for different $k$ with fixed $n = 0.75$ 
and $a_0 = 0.5$ \\

For different values of $k$ and $n$, we can generate a class of models of the universe in Bianchi type-I space-time 
with time dependent gravitational and cosmological constants. We observe that for $n \geq 2$ and $k = 1$, we obtain 
a class of transit models of the universe from early decelerated to present accelerating phase. For $n \leq 1$ and $k = 1$, 
we obtain accelerating models at present epoch. For examples: \\

\begin{itemize}
\item If we put $n = 2$ in Eq. (\ref{eq34}), we obtain $a(t) = \sqrt{(t^{k}e^{t})}$. In this case, we obtain the expressions 
for different physical and geometric quantities as obtained by Pradhan et al. \cite{ref72}. Thus, our investigations generalize 
the recent results of Pradhan et al. \cite{ref52}.

\item If we put $n = 2$ and $k = 1$ in Eq. (\ref{eq34}), we obtain $a(t) = \sqrt{(t e^{t})}$. In this case, we obtain the 
expressions for different physical parameters and geometric quantities by putting $n = 2$ and $k = 1$ in Eqs. 
(\ref{eq42})$-$(\ref{eq57}). Fig. $12$ depicts the variation of deceleration parameter with cosmic time for different values 
of ($n$, $k$). From Fig. $12$, we observe that for $n =2$ and $k = 1$ the model has a transition from very early decelerated 
phase to the present accelerating phase. We have already mentioned in Sect. $3$ that in such type of universe, the deceleration 
parameter must show signature interchange \cite{ref72}. The variation of energy density, cosmological constant and density 
parameters with cosmic time $t$ have been examined and found to have similar character like model given by Eq. (\ref{eq41}) 
in prevous section. 

\item If we put $n = 3$ and $k = 1$ in Eq. (\ref{eq34}), we obtain $a(t) = (t e^{t})^{\frac{1}{3}}$. In this case, we obtain the 
expressions for different physical parameters and geometric quantities as usual. From Fig. $12$, we observe that for $n =3$ and 
$k = 1$, the model has transition from early decelerated phase to the present accelerating phase. The variation of energy density, 
cosmological constant and density parameters with cosmic time $t$ have been examined and found the same as in previous case. 
The present value $q_{0}$ of the deceleration parameter obtained from observations are $ -1.27 \leq q_{0} \leq 2$ \cite{ref73}. 
Studies of galaxy from redshift surveys provide a value of $q_{0} = 0.1$, with an upper limit of $q_{0} < 0.75$ \cite{ref73}. 
Recent observations show that the deceleration parameter of the universe is in the range $ - 1 \leq q \leq 0$ i.e $q_{0} \approx 
-0.77$. First, we set $n = 3$ and $k = 1$ in Eq. (\ref{eq48}), we obtain $q_{0} = -0.67$. This value is very near to the  observed 
value of deceleration parameter (i.e., $q_{0} \approx -0.77$) at present epoch \cite{ref57}. Secondly, if we choose $n = 3$ and 
$k = 1$, we observe that all the values of physical and geometric parameters are easily integrable. Hence this case is important 
from physical aspects.

\item If we put $n = 1$ and $k = 1$ in Eq. (\ref{eq34}), we obtain $a(t) = t e^{t}$. In this case, we obtain the expressions for 
different physical parameters and geometric quantities as usual. From Fig. $12$, one can see that for $n = 1$ and $k = 1$ the model 
is accelerating at present epoch. The other physical parameters have the same property as already discussed.

\item If we put $k = 0$ in Eq. (\ref{eq34}), we obtain $a(t) = e^{t/n}$. In this case the universe has non-singular origin which 
seems reasonable to envision the dynamics of future universe. In this case, we found that energy density ($\rho$) is always negative 
and hence it is an unphysical case. We plan to work out a physically viable non-singular model in forthcoming paper.

\item We have also plotted $H(z)/(1 + z)$ data (32 points) and model prediction as a function of $z$ for different values of 
$k$ and $n$ by using the data given by Farooq and Ratra \cite{ref69}.
\end{itemize}

\section{Conclusions}
In this paper, we have presented a new class of models of accelerating universe and transit universe with gravitational 
coupling $G(t)$ and cosmological term $\Lambda(t)$ in the framework of general relativity. The models represent 
expanding, shearing and non-rotating universe. The parameters $H$, $\theta$, and $\sigma$ diverge at the initial 
singularity. There is a Point Type singularity \cite{ref58} at $t = 0$ in the models. The rate of expansion slows down 
and finally tends to zero at $t \to 0$. The pressure, energy density and cosmological term $\Lambda$ become negligible 
whereas the scale factors, gravitational constant $G$ and spatial volume become infinite as $t \to \infty$. The nature 
of decaying vacuum energy density $\Lambda(t)$ in our derived models is supported by recent cosmological observations. 
We observe that our derived models are isotropic at present epoch which is in good agreement with the current observations.\\

For different choice of $n$ and $k$, we can generate a class of viable cosmological models of the universe in Bianchi type 
space-time as well as in FRW universe. For example: if we set $n = 2$ in Eq. (\ref{eq34}), we find $a = \sqrt{t^{k}e^{t}}$ which 
is used by Pradhan and Amirhashchi \cite{ref55} in studying the accelerating dark energy models in Bianchi type-V space-time 
and Pradhan et al. \cite{ref74} in studying Bianchi type-I in scalar-tensor theory of gravitation. If we set $k = 1, ~ n= 2$ 
in  Eq. (\ref{eq34}), we find $a = \sqrt{t e^{t}}$ which is utilized by Amirhashchi et al. \cite{ref75} in studying 
interacting two-fluid scenario for dark energy in FRW universe. If we set $k = 1, ~ n= 1$ in  Eq. (\ref{eq34}), we find 
$a = t e^{t}$ which is exercised by Pradhan et al. \cite{ref76} to study the dark energy model in Bianchi type-$VI_{0}$ 
universe. It is observed that such models are also in good harmony with current observations. The present work generalizes 
the recent works (Pradhan et al. \cite{ref20,ref52}). 

\section*{Acknowledgments}
A. Pradhan would like to thank the Laboratory of Information Technologies, Joint Institute for Nuclear Research, Dubna, Russia 
for providing facility and supports where this work is done. A. Pradhan also acknowledges conscientiously the support in part by 
the University Grants Commission, New Delhi, India under the grant Project F.No. 41-899/2012 (SR). This work is also supported 
in part by a joint Romanian-LIT, JINR, Dubna, Research Project Theme No. 05-6-1060-2005/3013. 

\end{document}